\documentclass[aps,showpacs,superscriptadress,preprint]{revtex4}
\usepackage{graphicx}

\begin{document}
\title{The transition of equation of state of effective dark energy in the DGP model with bulk contents}

\author{Shaoyu Yin, Bin Wang}
\email{wangb@fudan.edu.cn} \affiliation{Department of Physics,
Fudan University, 200433 Shanghai}
\author{Elcio Abdalla}
\email{eabdalla@fma.if.usp.br} \affiliation{Instituto de Fisica,
Universidade de Sao Paulo, C.P.66.318, CEP 05315-970, Sao Paulo}
\author{Chi-Yong Lin}
\email{lcyong@mail.ndhu.edu.tw} \affiliation{Department of
Physics, National Dong Hwa University, Shoufeng, 974 Hualien}

\begin{abstract}
We investigate the effect of the bulk contents in the DGP braneworld
on the evolution of the universe. We find that although the pure DGP
model cannot accommodate the transition of the effective equation of
state of dark energy, once the bulk matter $T^5_5$ is considered,
the modified model can realize the $w_{eff}$ crossing $-1$. However
this transition of the equation of state cannot be realized by just
considering bulk-brane energy exchange or the GB effect while the
bulk matter contribution is not included. $T^5_5$ plays the major
role in the modified DGP model to have the $w_{eff}$ crossing $-1$
behavior. We show that our model can describe the super-acceleration
of our universe with the equation of state of the effective dark
energy and the Hubble parameter in agreement with observations.
\end{abstract}


\maketitle

\section{Introduction}

The accelerated expansion of our universe is one of the most
important discovery in the last decade
\cite{Riess:1998,Perlmutter:1999,Riess:2004}, having triggered
plenty of efforts to understand and explain it. This phenomenon is
in conflict with our common sense about attractive gravity. Within
the framework of general relativity, the acceleration is
attributed to the mysterious ``dark energy" existing in our
universe. The theoretical nature and origin of this dark energy
are a source of much debate. Candidates suggested for this dark
energy can be classified according to the behavior of their
respective equation of state $w=P/\rho$. The cosmological
constant, with $w =-1$, is located at a central position among
dark energy models both in theoretical investigation and in data
analysis \cite{Weinberg:1989}. In quintessence \cite{Caldwell:1998
et al}, Chaplygin gas \cite{Kamenshchik:2001} and holographic dark
energy models \cite{Li:2004}, $w$ always remains bigger than $-1$.
The phantom models of dark energy have $w <-1$ \cite{Caldwell et
al}. Recent more accurate data analysis tells us a dramatic
result, namely that the time varying dark energy gives a better
fit than a cosmological constant and in particular, $w$ can cross
$-1$ around $z = 0.2$ from above to below \cite{Alam et al}.
Theoretical attempts towards the understanding of the $w$ crossing
$-1$ phenomenon have been suggested, including the model
containing a negative kinetic scalar field and a normal scalar
field \cite{Feng et al}, a single scalar field model
\cite{Li:2005}, interacting holographic dark energy models
\cite{Wang et al} and others \cite{Nojiri et al}.

An alternative approach which does not need dark energy to explain
the late-time acceleration is motivated by string theory via the
brane-world scenarios. In this scenario our universe is a 3-d
brane embedded in a space-time with extra dimensions. The
cosmological evolution on the brane is described by an effective
Friedmann equation incorporating non-trivially with the effects of
the bulk onto the brane. The presence of the 5-d matter can
interact with the matter contents on the brane and alter the
cosmic expansion leading to a behavior resembling the dark energy.
The cosmic evolution of the Randall-Sundrum(RS) braneworld
\cite{Randall:1999} with energy exchange between brane and bulk
has been studied \cite{Kirstsis et
al,Cai:2006,Apostolopoulos:2006,Bogdanos:2006two,Sheykhi:2007}. In
these models, due to the energy exchange between the bulk and the
brane, the usual energy conservation law on the brane is broken
and consequently it was found that the equation of state of the
effective dark energy can experience the transition behavior
\cite{Cai:2006,Apostolopoulos:2006,Bogdanos:2006two,Sheykhi:2007}.

In string theory, in addition to the Einstein action, some higher
derivative curvature terms have been included to derive gravity. The
combination of the Einstein-Hilbert and Gauss-Bonnet(GB) term
constitutes, for 5D spacetimes, the most general Lagrangian to
produce second-order field equations
\cite{Zwiebach&Boulware,Lovelock:1971}. The GB correction changes
the bulk field equations and modifies the braneworld Friedmann
equation. It influences the evolution of the universe in our brane.
Effects of the GB correction on the RS braneworld have been studied
in \cite{Kofinas:2003,Sheykhi:2007}.

In this paper we are going to concentrate on another braneworld
model introduced by Dvali, Gabadadze and Porrati (DGP)
\cite{Dvali:2000}, where the braneworld is embedded in the flat bulk
with infinite extra dimensions. Considering that the graviton
propagates into the extra dimension, and at large scale, gravity can
become weaker due to its leakage, the DGP model can realize the
accelerated expansion naturally. However for the pure DGP model, its
effective equation of state never goes down to the phantom phase.
Our main motivation here is to investigate the effects of the bulk
contents in the DGP braneworld on the evolution of the universe and
explore the possibility of the transition of equation of state if
there are contributions from the bulk-related energy-momentum tensor
components which has been observed in RS model\cite{Kirstsis et
al,Cai:2006,Apostolopoulos:2006,Bogdanos:2006two,Sheykhi:2007}. The
DGP model only with $T^0_5$ has been investigated in
\cite{Kofinas:2006}. We are going to present a systematic and
complete examination of the bulk effects including $T^0_5$ and
$T^5_5$ terms on DGP model. Besides we will also study the
modification on the brane evolution due to the GB correction
together with bulk related energy-momentum tensor components.
Influences of the GB correction on the pure DGP braneworld have been
studied in \cite{Brown:2006,Cai:2005}. Although the effects of the
GB correction term on the late time universe is small, we will see
that it still plays an important role in the early time cosmic
evolution. We will show that in the DGP model the bulk matter
contribution $T^5_5$ plays a major role in accommodating the
transition of equation of state, while the $T^0_5$ and GB correction
alone cannot present a profile of the $w_{eff}$ crossing $-1$
phenomenon found by observations.

The organization of the paper is the following: in section II we
will give out the basic equation sets for the DGP model by
considering different correction terms respectively. The bulk
effects due to $T^5_5$ term will be shown in detail in section
III. In section IV, we will consider the influence of the energy
flow $T^0_5$ on the brane universe evolution. Conclusions and
discussions will be presented in the last section.

\section{General equations for DGP model with GB correction}

The DGP brane model with GB correction starts from the action
\begin{equation}
S=-\frac{1}{2\kappa^2}\int d^5X\sqrt{-g}(R-2\Lambda_5+\alpha L_{GB})
-\frac{1}{2\mu^2}\int
d^4x\sqrt{-\widetilde{g}}(\widetilde{R}-2\Lambda_4)+\int
d^5\sqrt{-g}L_{EM},
\end{equation}
where $\kappa$ and $\mu$ are related to the gravitational
constants and the Planck masses for the bulk and brane as
\cite{Deffayet:2001}:
\begin{equation}
\kappa^2=8\pi G_{(5)}=M_5^{-3};\qquad\mu^2=8\pi G_{(4)}=M_4^{-2},
\end{equation}
respectively, $\Lambda_5$ and $\Lambda_4$ are cosmological
constants for the bulk and brane. $L_{EM}$ is the energy-momentum
tensor and $L_{GB}$ is the GB correction term in the form
\begin{equation}
L_{GB}=R^2-4R^{AB}R_{AB}+R^{ABCD}R_{ABCD}.
\end{equation}
$\alpha$ is the coefficient of the GB term, which is positive, as
required by string theory \cite{Zwiebach&Boulware} and is
generally considered to be very small. If we take $\alpha=0$,
Eq.(1) reproduces the pure DGP model \cite{Deffayet:2001}.
Throughout the paper the capital letter are used to present the
5-d indices, while the Greek alphabet is used for 4-d brane case.

From the action one can obtain the field equation
\begin{equation}
G_{AB}+\Lambda_5g_{AB}+2\alpha
H_{AB}=\kappa^2\{T_{AB}-[\frac{1}{\mu^2}(\widetilde{G}_{\mu\nu}+
\Lambda_4\widetilde{g}_{\mu\nu})+\widetilde{T}_{\mu\nu}]\delta(y_b)\delta_A^{\mu}\delta_B^{\nu}\},
\end{equation}
where
$H_{AB}=RR_{AB}-2R^C_AR_{BC}-2R^{CD}R_{ACBD}+R^{CDE}_AR_{BCDE}-\frac{1}{4}g_{AB}L_{GB}$
is the second-order Lovelock tensor \cite{Lovelock:1971},
$\delta(y_b)$ comes from the difference between the integration
with 4-d metric and 5-d metric.

The energy-momentum tensor on the brane is assumed to be that of a
perfect fluid,
\begin{equation}
\widetilde{T}_{\mu\nu}=(\rho+p)u_{\mu}u_{\nu}+p\tilde{g}_{\mu\nu},
\end{equation}
where $u_{\mu}$, $\rho$ and $p$ are the fluid velocity, energy
density and pressure, respectively ($c=1$ is used). The non-zero
components related to the fifth dimension in the bulk
energy-momentum tensor are supposed to be $T^0_5$ and $T^5_5$,
whose role in the accelerated expansion will be studied in detail.

Generally, the metric in 5-d brane cosmology is written as
\begin{equation}
ds^2=-n^2(t,y)dt^2+a^2(t,y)\gamma^{ij}dx_idx_j+b^2(t,y)dy^2,
\end{equation}
where $y$ stands for the extra dimension orthogonal to the brane,
and $\gamma^{ij}$ is the maximally symmetric 3-d tensor. Then
$\sqrt{-g}=b\sqrt{-\widetilde{g}}$, thus
$\delta(y_b)=\frac{\delta(y)}{b}$. From this metric the Einstein
equation can be obtained directly. According to Eq.(4) one can
obtain the Einstein tensor components as \cite{Barcelo:2003}
\begin{eqnarray}
G_{tt}&=&3[n^2\Phi+\frac{\dot{a}}{a}\frac{\dot{b}}{b}-\frac{n^2}{b^2}(\frac{a''}{a}-\frac{a'}{a}\frac{b'}{b})],\nonumber\\
G_{ty}&=&3(\frac{\dot{a}}{a}\frac{n'}{n}+\frac{a'}{a}\frac{\dot{b}}{b}-\frac{\dot{a}'}{a}),\nonumber\\
G_{ij}&=&\frac{a^2}{b^2}\gamma_{ij}[\frac{a'}{a}(\frac{a'}{a}+2\frac{n'}{n})
-\frac{b'}{b}(\frac{n'}{n}+2\frac{a'}{a})+2\frac{a''}{a}+\frac{n''}{n}]\nonumber\\
&&-\frac{a^2}{n^2}\gamma_{ij}[\frac{\dot{a}}{a}(\frac{\dot{a}}{a}-2\frac{\dot{n}}{n})
-\frac{\dot{b}}{b}(\frac{\dot{n}}{n}-2\frac{\dot{a}}{a})+2\frac{\ddot{a}}{a}+\frac{\ddot{b}}{b}]-k\gamma_{ij},\nonumber\\
G_{yy}&=&3[-b^2\Phi+\frac{a'}{a}\frac{n'}{n}-\frac{b^2}{n^2}(\frac{\ddot{a}}{a}-\frac{\dot{a}}{a}\frac{\dot{n}}{n})],\nonumber\\
H_{tt}&=&6\Phi[\frac{\dot{a}}{a}\frac{\dot{n}}{n}+\frac{n^2}{b^2}(\frac{a'}{a}\frac{b'}{b}-\frac{a''}{a})],\nonumber\\
H_{ty}&=&6\Phi(\frac{\dot{a}}{a}\frac{n'}{n}+\frac{a'}{a}\frac{\dot{b}}{b}-\frac{\dot{a}'}{a}),\nonumber\\
H_{ij}&=&2a^2\gamma_{ij}\{\Phi[\frac{1}{n^2}(\frac{\dot{n}}{n}\frac{\dot{b}}{b}-\frac{\ddot{b}}{b})
-\frac{1}{b^2}(\frac{n'}{n}\frac{b'}{b}-\frac{n''}{n})]\nonumber\\
&+&\frac{2}{a^2bn}[\frac{\dot{a}^2\dot{b}\dot{n}}{n^4}
+\frac{a'^2b'n'}{b^4}+\frac{\dot{a}a'}{b^2n^2}(b'\dot{n}-\dot{b}n')]\nonumber\\
&-&2[\frac{1}{n^2}\frac{\ddot{a}}{a}(\frac{1}{n^2}\frac{\dot{a}}{a}\frac{\dot{b}}{b}+\frac{1}{b^2}\frac{a'}{a}\frac{b'}{b})
+\frac{1}{b^2}\frac{a''}{a}(\frac{1}{n^2}\frac{\dot{a}}{a}\frac{\dot{n}}{n}+\frac{1}{b^2}\frac{a'}{a}\frac{n'}{n})]\nonumber\\
&+&\frac{2}{b^2n^2}[\frac{\ddot{a}}{a}\frac{a''}{a}-\frac{\dot{a}^2}{a^2}\frac{n'^2}{n^2}-
\frac{a'^2}{a^2}\frac{\dot{b}^2}{b^2}-\frac{\dot{a}'}{a}(\frac{\dot{a}'}{a}-2\frac{\dot{a}}{a}\frac{n'}{n}-
2\frac{a'}{a}\frac{\dot{b}}{b})]\},\nonumber\\
H_{yy}&=&6\Phi[\frac{a'}{a}\frac{n'}{n}+\frac{b^2}{n^2}(\frac{\dot{a}}{a}\frac{\dot{n}}{n}-\frac{\ddot{a}}{a})],
\end{eqnarray}
where
\begin{equation}
\Phi=\frac{1}{n^2}\frac{\dot{a}^2}{a^2}-\frac{1}{b^2}\frac{a'^2}{a^2}+\frac{k}{a^2}.
\end{equation}
The dot denotes a derivative with respect to $t$, and the prime
the derivative with respect to $y$. Without loosing generality, in
the brane world scenario, one usually chooses the metric function
$b(t,y)=1$ and $n(t,0)=1$ to simplify the calculation.

We choose the brane to be located at $y=0$, and suppose the metric
functions to be continuous at this point, but their first
derivatives are discontinuous due to the energy-momentum tensor
distribution on the brane. Furthermore, the geometry is supposed to
display a $Z_2$-symmetry around $y=0$, thus $a'(0_+)=-a'(0_-)$ and
$n'(0_+)=-n'(0_-)$. If one integrates the field equation for the
$tt$ and $ij$ components at the infinitesimal region near $y=0$,
only those terms in the metric with $a''$ or $n''$ and the
energy-momentum distribution on the brane can remain. Then the
differences of $a'$ and $n'$ on both sides of the brane, say,
$a'(0_+)-a'(0_-)\equiv2a'(0_+)$ and $n'(0_+)-n'(0_-)\equiv2n'(0_+)$
can be obtained. For simplicity, throughout the paper we use $a'$
and $n'$ to stand for $a'(0_+)$ and $n'(0_+)$, and in the equation
on the brane all bulk terms are taken with their values at $y=0_+$.
From $G_{tt}$, $G_{ij}$, $H_{tt}$, $H_{ij}$ in Eq.(7), we find that
$a'$ and $n'$ satisfy the following equations:
\begin{equation}
\frac{a'}{a}\{-3+4\alpha[\frac{a'^2}{a^2}-3(\frac{k}{a^2}+\frac{\dot{a}^2}{a^2})]\}
=\frac{\kappa^2}{2\mu^2}[\mu^2\rho-3(\frac{k}{a^2}+\frac{\dot{a}^2}{a^2})];
\end{equation}
\begin{equation}
n'+\frac{2a'}{a}+
4\alpha[n'(\frac{k}{a^2}-\frac{a'^2}{a^2}+\frac{\dot{a}^2}{a^2})
+\frac{2a'}{a}(\frac{\ddot{a}}{a}-\frac{\dot{a}\dot{n}}{a})]
=\frac{\kappa^2}{2\mu^2}(\mu^2p+\frac{k}{a^2}+\frac{\dot{a}^2}{a^2}
-\frac{2\dot{a}\dot{n}}{a}+\frac{2\ddot{a}}{a}).
\end{equation}

Generally, one can solve these equations and substitute the results
of $a'$ and $n'$ into the field equation for the $ty$ and $yy$
components to obtain the continuity equation and the effective
Friedmann equation. But before that, it is helpful to notice that
the function $\Phi$ we introduced in Eq.(8) satisfies
\begin{eqnarray}
\widetilde{\Phi}'&=&\frac{2\kappa^2}{3}(\Lambda_5-T^0_0)a^3a'-\frac{2\kappa^2}{3}T^0_5a^3\dot{a},\\
\dot{\widetilde{\Phi}}&=&\frac{2\kappa^2}{3}(\Lambda_5-T^5_5)a^3\dot{a}-\frac{2\kappa^2}{3}\frac{n^2}{b^2}T^0_5a^3a',
\end{eqnarray}
where $\widetilde{\Phi}\equiv(\Phi+2\alpha\Phi^2)a^4$. From (12), if
$T^0_5=0$ and $T^5_5$ has a proper \textit{ansatz}, such as
$T^5_5=\frac{F}{\kappa^2}a^{\nu}$, $\widetilde{\Phi}$ can be
obtained analytically by an integration with respect to $t$:
\begin{equation}
\widetilde{\Phi}=\frac{\kappa^2}{6}\Lambda_5a^4-\frac{2Fa^{\nu+4}}{3(4+\nu)}+C,
\end{equation}
where $C$ is an integration constant. For the case without GB
term, the solution of $\Phi$ is simply
\begin{equation}
\Phi=\frac{\kappa^2}{6}\Lambda_5-\frac{2Fa^{\nu}}{3(4+\nu)}+\frac{C}{a^4},
\end{equation}
where the term $\frac{C}{a^4}$ is usually referred to the dark
radiation \cite{Maartens:2000}. For the case with GB correction,
solutions of $\Phi$ are,
\begin{equation}
\Phi=\frac{-(4+\nu)\pm\sqrt{(4+\nu)^2}\sqrt{1+8\alpha(\frac{\kappa^2\Lambda_5}{6}+\frac{C}{a^4}
-\frac{2Fa^{\nu}}{3(4+\nu)})}}{4\alpha(4+\nu)},
\end{equation}
while only one solution with finite $\alpha\rightarrow0$ limit can
be taken. For $\nu>-4$ the solution reads
\begin{equation}
\Phi=\frac{-1+\sqrt{1+8\alpha(\frac{\kappa^2\Lambda_5}{6}+\frac{C}{a^4}
-\frac{2Fa^{\nu}}{3(4+\nu)})}}{4\alpha}.
\end{equation}
When $\alpha\rightarrow0$, Eq.(16) goes back to Eq.(14).

From the definition of $\Phi$ in Eq.(8), we see that $a'$ can be
expressed in terms of $\dot{a}$ once $\Phi$ is obtained.
Integrating the equation of $tt$ component around $y=0$ and
substituting $a'$ in terms of $\Phi$ and $\dot{a}$, we can finally
arrive at the equation for the Hubble parameter
$H(t)\equiv\frac{\dot{a}}{a}$,
\begin{equation}
(H^2+\frac{k}{a^2}-\Phi)[1+\frac{8\alpha}{3}(H^2+\frac{k}{a^2}+\frac{\Phi}{2})]^2=
\frac{r^2}{4}[H^2+\frac{k}{a^2}-\frac{\mu^2}{3}(\rho+\Lambda_4)]^2,
\end{equation}
where $r=\kappa^2/\mu^2$ is the DGP crossover radius.

For the sake of simplicity and clarity in the following
discussion, we give the simplifications we are going to use. We
will neglect the cosmological constant, $\Lambda_5=\Lambda_4=0$,
since the effect of the cosmological constant can be included in
$\rho$ and $p$. We will apply our discussion to the flat universe
with $k=0$. Besides, we will employ dimensionless notations in the
following calculation by defining
\begin{eqnarray}
x&=&\frac{H^2}{H_0^2},\nonumber\\
z&=&\frac{a_0}{a}-1,\nonumber\\
u&=&\frac{\mu^2\rho}{3H_0^2},\nonumber\\
y&=&\frac{\Phi}{2H_0^2},\nonumber\\
n&=&\frac{1}{H_0^2r^2},\nonumber\\
m&=&\frac{8H_0^2}{3}\alpha,\nonumber\\
X&=&\frac{a_0^2}{H_0^2(4+\nu)}F,\nonumber\\
\widetilde{X}&=&\frac{a_0^2}{H_0^2}F,\nonumber\\
M&=&\frac{3}{2H_0^2a_0^4}C,
\end{eqnarray}
where $a_0$ and $H_0$ are the present values of the scale factor
and the Hubble parameter, $z$ is the redshift and
$u/x=\frac{\mu^2\rho}{3H^2}$ is the proportion of matter in the
total effective energy density.

Using dimensionless notations, the expression for the solution of
$\Phi$ becomes
\begin{equation}
\Phi=2H_0\frac{-X(1+z)^{-\nu}+M(1+z)^4}{3},
\end{equation}
for the case without the GB correction; if the GB term is
included, it reads
\begin{equation}
\Phi=2H_0\frac{-1+\sqrt{1+2m(-X(1+z)^{-\nu}+M(1+z)^4)}}{3m}.
\end{equation}
The equation for $H^2$, Eq.(17), turns into an equation for $x$
\cite{Cai:2005}
\begin{equation}
4n(x-2y)[1+m(x+y)]^2=(x-u)^2.
\end{equation}

If $T^0_5$ is nonzero, $\Phi$ cannot be solved analytically and we
can not take advantage of the simplicity discussed above. To obtain
the equation for $H^2$, we need to substitute the solutions of $a'$
and $n'$ into the equation of $yy$ component, and obtain $H(t)$
through onerous calculations. The acceleration of the scale factor
can be written as $\ddot{a}\equiv a(H^2+\dot{H})$. By choosing the
proper \textit{ansatz} of $T^0_5$ and expressing the result of
$\rho(t)$ as a function of $a(t)$, we can generally obtain the
equation as a nonlinear ordinary differential equation of $H(t)$,
combining with the unknown function $a(t)$. To solve such a problem
in RS model \cite{Cai:2006,Bogdanos:2006two} the authors introduced
new effective fields related to $H(t)^2$ and separated the equation
into two equations, both of which are solvable separately. But for
the DGP model, due to the extra 4-d intrinsic curvature terms, the
highest order of $H(t)$ is $4$, rather than $2$ in the RS model.
With the GB correction, the order goes up to $8$. In
Ref.\cite{Kofinas:2006} when just nonzero $T^0_5$ was included in
the pure DGP model, the author solved the problem by introducing the
concept of ``fix point" and setting $\rho(t)$ and the auxiliary
field time-independent. Generally, we do not hope to obtain any
analytical solution for such a nonlinear ordinary differential
equation. We will count more on the numerical calculations.
Considering nonzero $T^0_5$ and $T^5_5$ components, our problem is
general and complicated. We will present a general way to solve the
problem.

We have two time-dependent functions, $H(t)$ and $a(t)$, in the same
equation. Considering that in the big-bang cosmology the flat
universe is expanding monotonically, $a(t)$ is a monotonic function
of $t$, and $H(t)$ can be written as $H(a)$. For the convenience we
will use the dimensionless redshift $z$ and write $H(t)$ as
\begin{equation}
\dot{H}(t)=-\frac{H_0^2(1+z)}{2}\frac{dx(z)}{dz}.
\end{equation}

Substituting all the dimensionless notations into the equation of
the $yy$ component and expressing the results until the linear
order in $\alpha$, the equation for $H(t)$, or equally, $x(z)$, is
\begin{eqnarray}
0&=&-mx^4+m[24n+u+(1+z)x']x^3\nonumber\\
&+&[-16n+48mn^2-12mnu+3mu^2-3m(6n+u)(1+z)x']x^2\nonumber\\
&+&\{64n^2+8nu-12mnu^2-5mu^3+[8n(1-3mn)+3mu(8n+u)](1+z)x'\}x\nonumber\\
&+&[8nu^2+2mu^4-\{16n^2+u[8n+mu(6n+u)]\}(1+z)x']\nonumber\\
&+&\frac{32n^2(1+z)^{-\nu}}{3}\widetilde{X},
\end{eqnarray}
where the prime here is the derivative with respect to $z$, $x$ and
$u$ are functions of $z$, and we have taken
$T^5_5=\frac{F}{\kappa^2}a^{\nu}$. It is to be noted that since we
don't need to  analytically  integrate $T^5_5$ with the term
$a^3\dot{a}$ as did in Eq.(12), we can  in principle use any form of
$T^5_5$ as a function of $a$. Eq.(23) is a nonlinear differential
equation of $x(z)$.

From the equation of the $ty$ component, assuming $b(t,y)=1$, we
get
\begin{equation}
3(1+4\alpha\Phi)(\frac{\dot{a}}{a}\frac{n'}{n}-\frac{\dot{a}'}{a})=T_{05}.
\end{equation}
Taking the value of each term in this equation at $y=0_+$ and
substituting the solution of $a'$ and $n'$, we can find that the
left hand side of this equation is simply
$\frac{1}{2}(\dot{\rho}+3H(\rho+p))$. If $T^0_5=0$, it is the
conservation of energy on the brane. If $T^0_5\neq0$, it acts as
the energy flow between the brane and the bulk,
\begin{equation}
\dot{\rho}+3H(\rho+p)=-2T^0_5.
\end{equation}
Here $T^0_5$ has a sign difference as compared to $T_{05}$ due to
the metric term $g_{tt}|_{y=0}=-n(t,0)=-1$.

Setting the \textit{ansatz}, $T^0_5=fHa^s$, the equation for $\rho$
can be solved analytically. Expressing  $\rho(t)$ as $\rho(a)$, we
have
\begin{equation}
\frac{d\rho}{da}+\frac{3(1+w)\rho}{a}+2fa^{s-1}=0,
\end{equation}
with the solution
\begin{equation}
\rho=a^{-3-3w}C_1-\frac{2fa^s}{3+3w+s},
\end{equation}
where $C_1$ is an integration constant. For the cold matter on the
brane $w=0$, the first term on the right-hand-side is proportional
to $a^{-3}$ and the second term could be attributed to the effective
dark energy. Eq.(27) can be expressed by using the dimensionless
notation $u(z)$,
\begin{equation}
u(z)=P(1+z)^{-s}+\Omega_{m0}(1+z)^3,
\end{equation}
where $P=-\frac{2\mu^2a_0^s}{3(3+s)H_0^2}f$, and
$\Omega_{m0}=\frac{\mu^2C_1}{3H_0^2a_0^3}=\frac{8\pi
G}{3H_0^2}\rho_0$ is the present ratio of conservative matter in the
total energy density of the universe, $\rho_0$ is the density of the
conservative matter today. There is a strong constraint on the value
of dimensionless parameter $P$. Since the matter portion of the
total energy density should be in the range $[0,1]$, thus
$0\leq\frac{u(z)}{x(z)}\leq1$. Here we will use the fitting results
on the WMAP data \cite{Spergel:2003} and take $\Omega_{m0}=0.28$ in
our calculation. Of course, that fitting is from a different model,
but the generally accepted values of $\Omega_{m0}$ are all close to
this value, and the small variation of this value will not change
the qualitative conclusion of our calculation. At the present moment
$z=0$, we have $-0.28\leq P\leq0.72$. Any solution with $P$ out of
this range is physically unreasonable.

To describe the effect of the effective dark energy, we can define
the effective equation of state \cite{Linder:2002}:
\begin{equation}
w(z)_{eff}\equiv-1+\frac{1}{3}\frac{d\ln\delta H^2}{d\ln(1+z)},
\end{equation}
where $\delta H^2\equiv H(z)^2-\Omega_{m0}(1+z)^3H_0^2$. $w_{eff}$
can be expressed by using the dimensionless parameters as
\begin{equation}
w(z)_{eff}=-1+\frac{(1+z)\frac{dx(z)}{dz}-3\Omega_{m0}(1+z)^3}{3x(z)-3\Omega_{m0}(1+z)^3}.
\end{equation}
The subscript $eff$ indicates that the effect similar to the dark
energy on the brane comes from the bulk contribution. Another
important quantity is the deceleration parameter $q$
\begin{equation}
q\equiv-\frac{\ddot{a}a}{\dot{a}^2}=\frac{1}{2}+\frac{3}{2}w(z)_{eff}(1-\frac{\Omega_{m0}(1+z)^3}{x(z)}),
\end{equation}
which will also be used in the following discussion of the
expansion of our universe.

\section{Calculation and Discussion without considering the $T^0_5$ term}

For the case without $T^0_5$ term, we can solve the equation Eq.(21)
by substituting $\Phi$ from Eq.(20) and $u$ from Eq.(28) but with
$P=0$. We can put the result into Eq.(30) and Eq.(31) to examine the
behavior of the effective dark energy.

In Eq.(21), if there is no GB correction, the equation for $x$ is
quadratic, thus we have two solutions. For the case $\Phi=0$, they
recover the two branches of pure DGP model, DGP(+) and DGP(-):
$x=2n+(1+z)^3\Omega_{m0}\pm2\sqrt{n^2+n(1+z)^3\Omega_{m0}}$. Since
only the DGP(+) solution has late-time self-accelerating behavior
\cite{Brown:2006}, we will concentrate our discussion on this
solution. When GB correction is added, Eq.(21) becomes a cubic
equation and has three roots for $x$, two of which correspond to
DGP(+) and DGP(-) in $\alpha\rightarrow0$ limit, and the third one
diverges when $\alpha\rightarrow0$. For comparison, we also study
the solution with DGP(+) limitation for that case in this work.

To compare our model description on the evolution of the universe
with the observation, we have several constraints to meet. At
present we have $x(z=0)\equiv H(t=0)^2/H_0^2\equiv1$. For the
effective equation of state, we require $w_{eff}(z=0.2)=-1$ and
$w_{eff}(z=0)=-1.06$ \cite{Alam et al}. We will use these
constraints to refine our model parameters. Assuming $T^0_5=0$, we
have parameters such as $n$ (corresponding to the DGP crossover
radius), $m$ (corresponding to the GB correction), $M$
(corresponding to the dark radiation), $X$ and $\nu$ (relating to
$T^5_5$ form). We will focus on whether we can accommodate the
$w_{eff}$ crossing $-1$ by including bulk related energy-momentum
tensor and the GB correction, which cannot be realized in the pure
DGP model.

Since we have the parameter $\nu$ in the exponential, the
equations are not polynomial. We will use \textit{FindRoot} in our
calculation which will raise the problem about the choice of the
initial values. To avoid the possibility of failing to find the
solution, we will try different initial values in the
\textit{FindRoot}. We find that the solution is not strongly
dependent on the initial values so that we are confident that our
results are almost the whole collection of all possible solutions
to the equations we deal with.

To see the consistency of our results with observation, we will plot
the Hubble parameter and compare with the observational $H(z)$ data
as shown in Table 1. It is interesting to note that all the cases
which can accommodate the equation of state transition can fit well
the observational $H(z)$ data. Remembering that the $w_{eff}$
crossing $-1$ was observed in the SNIa data fitting containing an
integration effect in the luminosity distance, while the Hubble
parameter does not suffer from this integrated over effect, the
Hubble parameter data can present a complementary and consistent
check for our model.

\begin{table}[b]
\begin{tabular}[t]{c|ccccccccc}
\hline\hline\centering
z & 0.09 & 0.17 & 0.27 & 0.40 & 0.88 & 1.30 & 1.43 & 1.53 & 1.75 \\
\hline
H(z) (km/s/Mpc)& 69 & 83 & 70 & 87 & 117 & 168 & 177 & 140 & 202 \\
1$\sigma$ uncertainty\quad&\quad$\pm12$\quad&\quad$\pm8.3$\quad&\quad$\pm14$\quad&\quad$\pm17.4$\quad&\quad$\pm23.4$\quad&\quad$\pm13.4$\quad&\quad$\pm14.2$\quad&\quad$\pm14$\quad&\quad$\pm40.4$\\
\hline\hline
\end{tabular}\label{table1}
\caption{The observational $H(z)$ data
\cite{Jimenez:2003,Simon:2005}.}
\end{table}

Now we list out our results step by step. In the following results,
all the numbers obtained in numerical calculation are expressed only
with 3 digits after decimal unless for the cases where more digits
are necessary to be shown.

1. DGP+M

In this step we consider the DGP model with the dark radiation,
where $M$ denotes the dark radiation. Now we have two free
parameters $n$ and $M$ and we will use two constraints: $x(0)=1$ and
$w_{eff}(0.2)=-1$ to see whether the dark radiation can help to
realize the $w_{eff}$ crossing $-1$. Actually $n$ and $M$ can be
solved by using these constraints as $n=0.157$ and $M=0.263$. But
using these values of $n$ and $M$, the $w_{eff}$ behavior is not
good. $w_{eff}$ crosses $-1$ at $z=0.2$ from below to up as shown in
Fig.1, and the present value is $w_{eff}=-0.950$. This is not in
consistent with the observation, especially the transition behavior.

\begin{figure}[!h]
\includegraphics[totalheight=10cm, width=10cm]{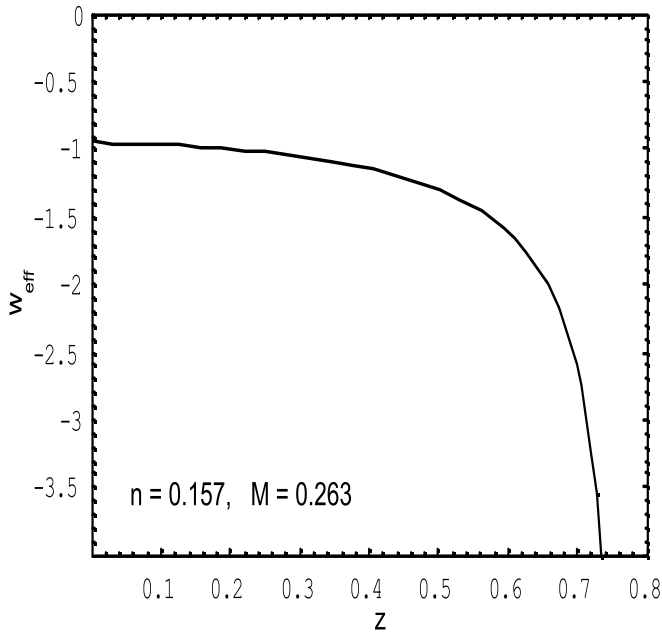}
\caption{The $w_{eff}$ curve as function of $z$ in the case DGP+M.
The behavior of $w_{eff}$ is bad since it crossed $-1$ at $z=0.2$
from below to above. } \label{fig1}
\end{figure}

2. DGP+GB+M

Including the GB correction, we have three free parameters now,
such as  $n$, $m$ and $M$. If we apply all three constraints
($x(0)=1$, $w_{eff}(0.2)=-1$ and $w_{eff}(0)=-1.06$), the solution
is complex ($n=-0.030+0.093i$, $m=-0.002-0.007i$ and
$M=6.340+6.244i$). If we apply only two constraints ($x(0)=1$ and
$w_{eff}(0.2)=-1$) and search $n$ in a big range $0.001\leq
n\leq5$, the similar result to that in case 1 appears: $w_{eff}$
crosses $-1$ from below to above and $m$ is negative. One solution
is shown in Fig.2a. We note that there is a singularity about
$z=1.246$ in the curve of $w_{eff}$. This singularity comes from
the definition in Eq.(29), we see that when $\delta H^2\equiv
H(z)^2-\Omega_{m0}(1+z)^3H_0^2\leq0$, the logarithm is not well
defined, but one can still calculate the $w_{eff}$ through the
simplified expression in Eq.(30). In Fig.2b we show the relation
between $H^2/H_0^2$ and the matter component $\Omega_{m0}(1+z)^3$.
We see that beyond the redshift $z=1.246$ the matter component is
overweight, so $\delta H^2<0$, which means the effective dark
energy component is negative. This is obviously unreasonable, and
at least it shows that the model fails in explaining the universe
before that redshift. In this work we concentrate on those
solutions with $w_{eff}(z)$ free of singularity.

\begin{figure}[!h]
\includegraphics[totalheight=8cm, width=8cm]{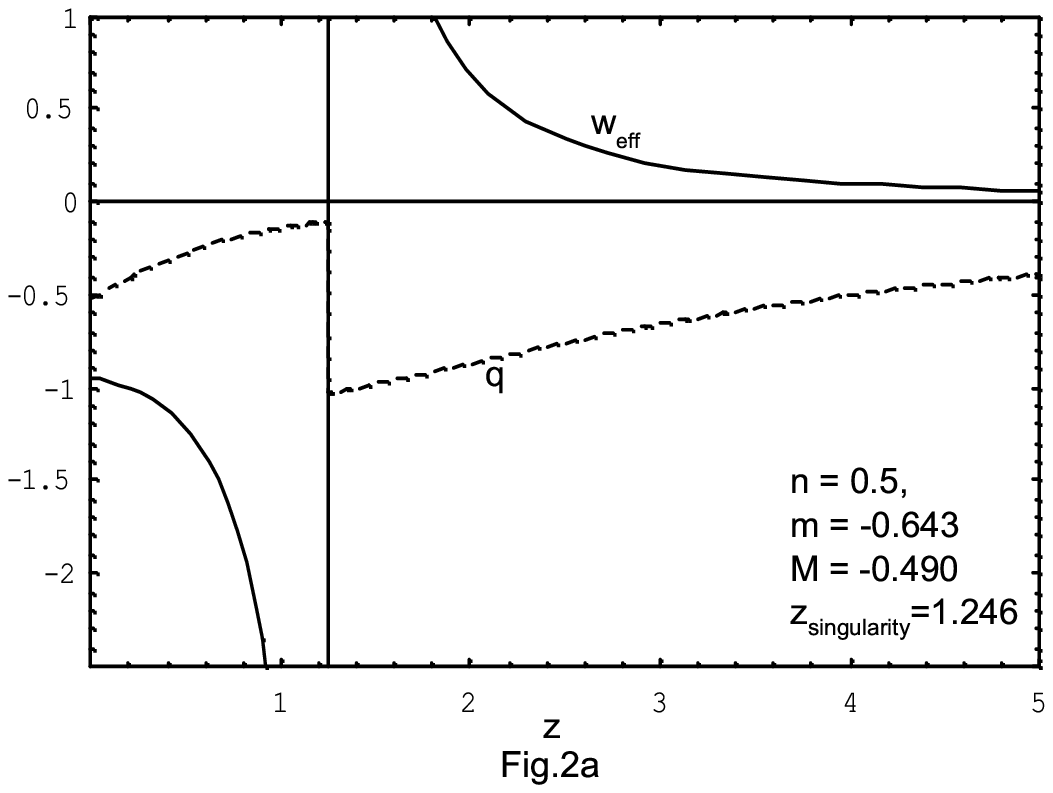}
\includegraphics[totalheight=8cm, width=8cm]{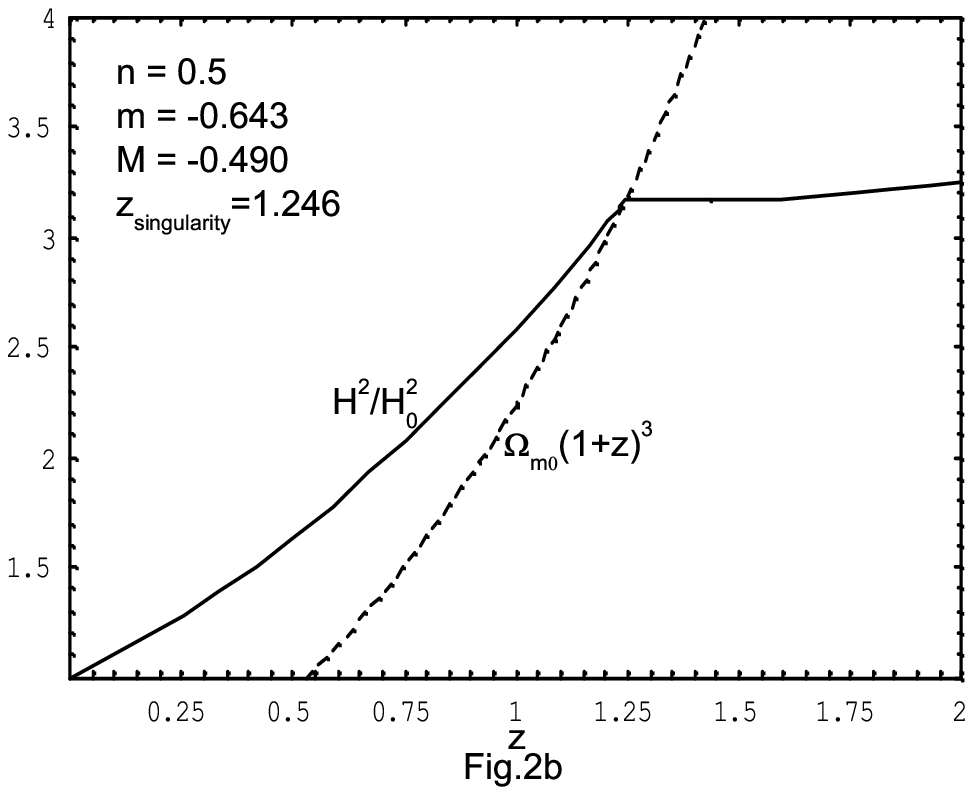}
\caption{In the case DGP+GB+M, $w_{eff}$ and $q$ as functions of $z$
are shown in Fig.2a. The behavior is not favored as in Fig.1.
Singularity is observed at $z=1.246$. In Fig.2b, relation between
$H^2/H_0^2$ and the matter component $\Omega_{m0}(1+z)^3$ is shown.
When $z>1.246$, $\delta H^2\equiv H(z)^2-\Omega_{m0}(1+z)^3H_0^2$
becomes negative, which breaks the definition of $w_{eff}$ in
Eq.(29).} \label{fig2}
\end{figure}

3. DGP+$T^5_5$

Now we include the bulk related energy-momentum tensor $T^5_5$. In
this case we have three free parameters ($n$, $X$ and $\nu$) and we
are going to employ all three constraints ($x(0)=1$,
$w_{eff}(0.2)=-1$ and $w_{eff}(0)=-1.06$). We can find the solution
$n=0.046$, $X=2.729$ and $\nu=0.948$. The curves of $w_{eff}$, $q$
and $H$ v.s. redshift $z$ are shown in Fig.3a and Fig.3b
respectively. In plotting Fig.3b, we have used $H_0=72km/s/Mpc$
\cite{Freemann:2000}. It is interesting to find that parameters
adjusted to meet the requirement of $w_{eff}$ crossing $-1$ and its
value at the present moment automatically fit well to the $H(z)$
data. Recalling that the transition behavior of $w_{eff}$ results
from the SN data analysis containing integration in the luminosity
distance, while the Hubble parameter is not integrated over, which
persists fine structure highly degenerated in the luminosity
distance, the simultaneous satisfaction of the $w_{eff}$ behavior
and the $H(z)$ data gives complementary and consistent check of the
viability of our model.

\begin{figure}[!h]
\includegraphics[totalheight=8cm, width=8cm]{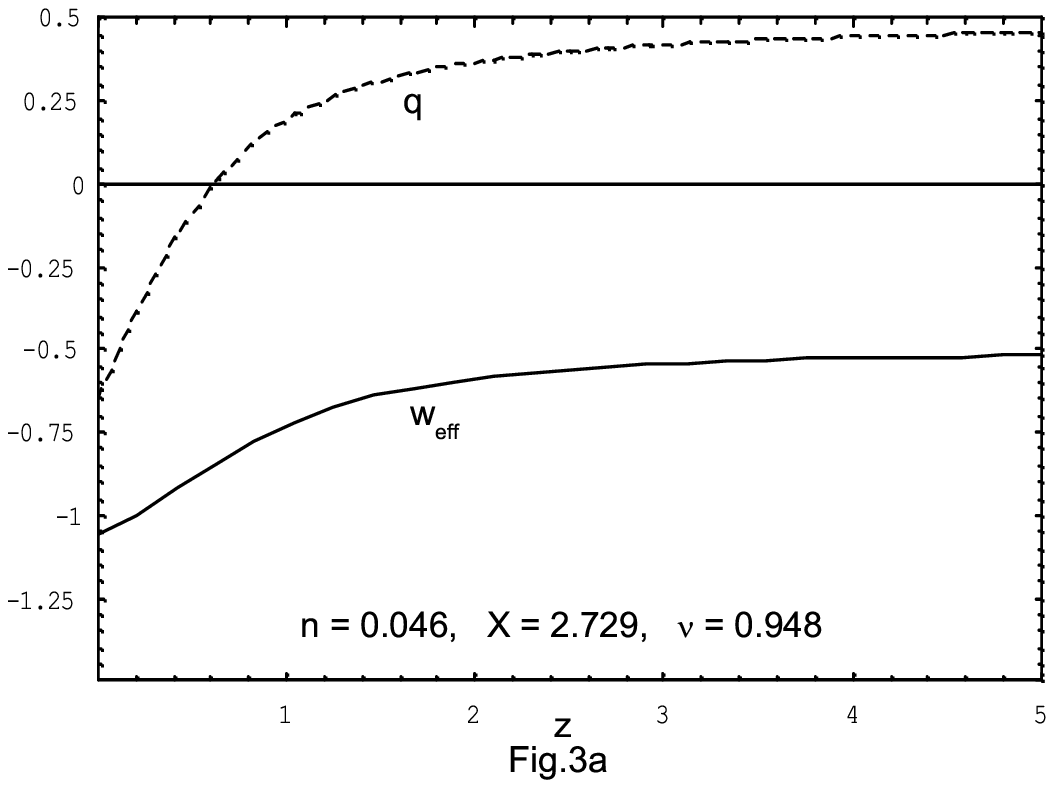}
\includegraphics[totalheight=8cm, width=8cm]{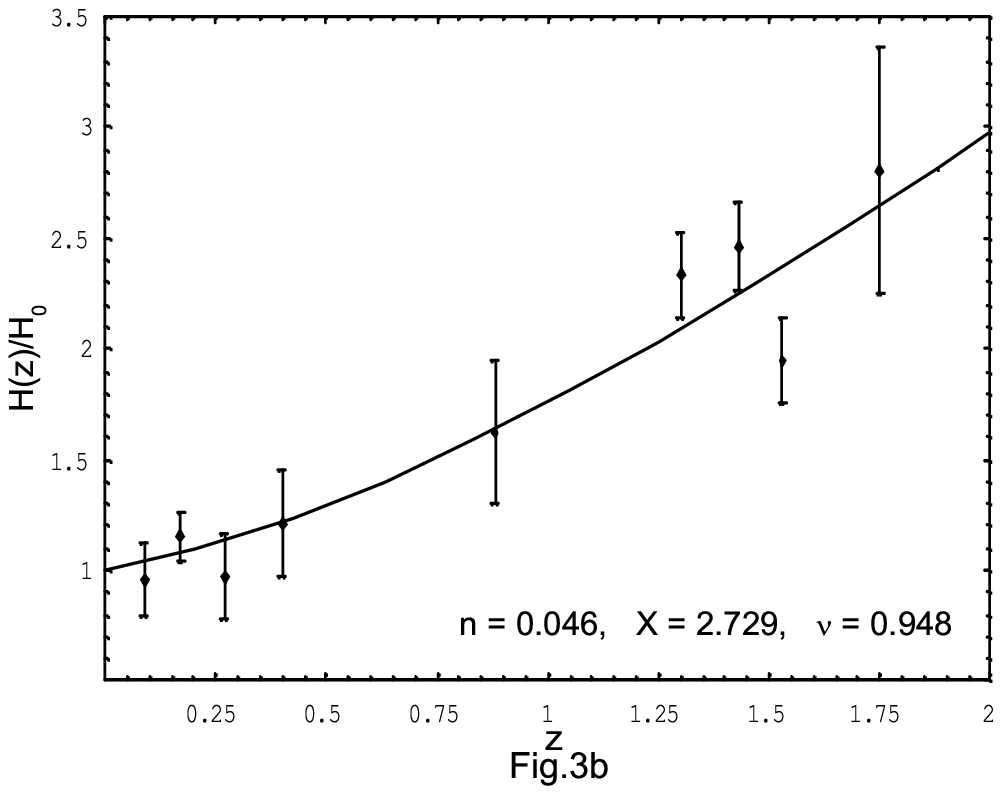}
\caption{$w_{eff}$ and $q$ v.s. $z$ (Fig.3a) and $H(z)$ curve
(Fig.3b) in DGP+$T^5_5$ case. We see that $H(z)$ curve fits the data
quite well.} \label{fig3}
\end{figure}

4. DGP+GB+$T^5_5$

Here we include the GB correction based on the case discussed above.
We now have four free parameters $n$, $m$, $X$ and $\nu$. Employing
constraints ($x(0)=1$, $w_{eff}(0.2)=-1$ and $w_{eff}(0)=-1.06$) and
searching through $0.001\leq n\leq5$, we see that $m$, $X$ and $\nu$
can either be negative or positive, but the latter two are always
with the same sign. $m$ decreases with the increase of $n$ and can
be positive only when $n<0.05$. The $w_{eff}$ curve has no
singularity for positive $m$, but contains singularity when $m<0$
(few solutions without singularity have been found with negative
$m$, but then $\nu$ becomes smaller than $-4$, which conflicts with
our simplification assumption discussed above). We are interested in
the positive $m$, since GB correction coefficient $\alpha$ is always
positive, the singularity-free curves of $w_{eff}(z)$ and $q(z)$
with positive solution $m=0.025$ are shown in Fig.4a. When the GB
correction is considered,  $w_{eff}$ appears more different at
larger $z$ if compared to the result without the GB correction.
Since the GB effect is only important in the early universe, its
stronger modification to the $w_{eff}$ at bigger redshift is
natural. In Fig.4b, we plotted the $H(z)$ curve by using the same
adjusted parameter from the constraints on the equation of state,
and we see again that in the model when the $w_{eff}$ requirement is
met, the $H$ parameter automatically fits the data, which gives the
consistent check of the model.

\begin{figure}[!h]
\includegraphics[totalheight=8cm, width=8cm]{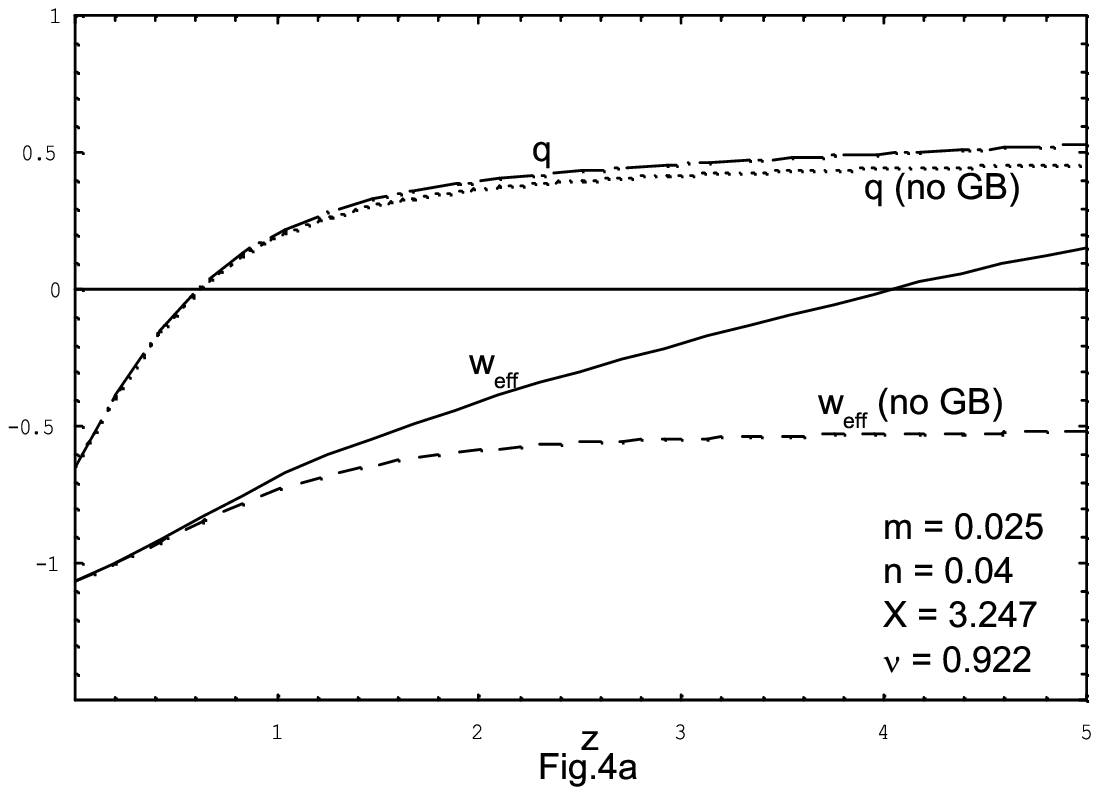}
\includegraphics[totalheight=8cm, width=8cm]{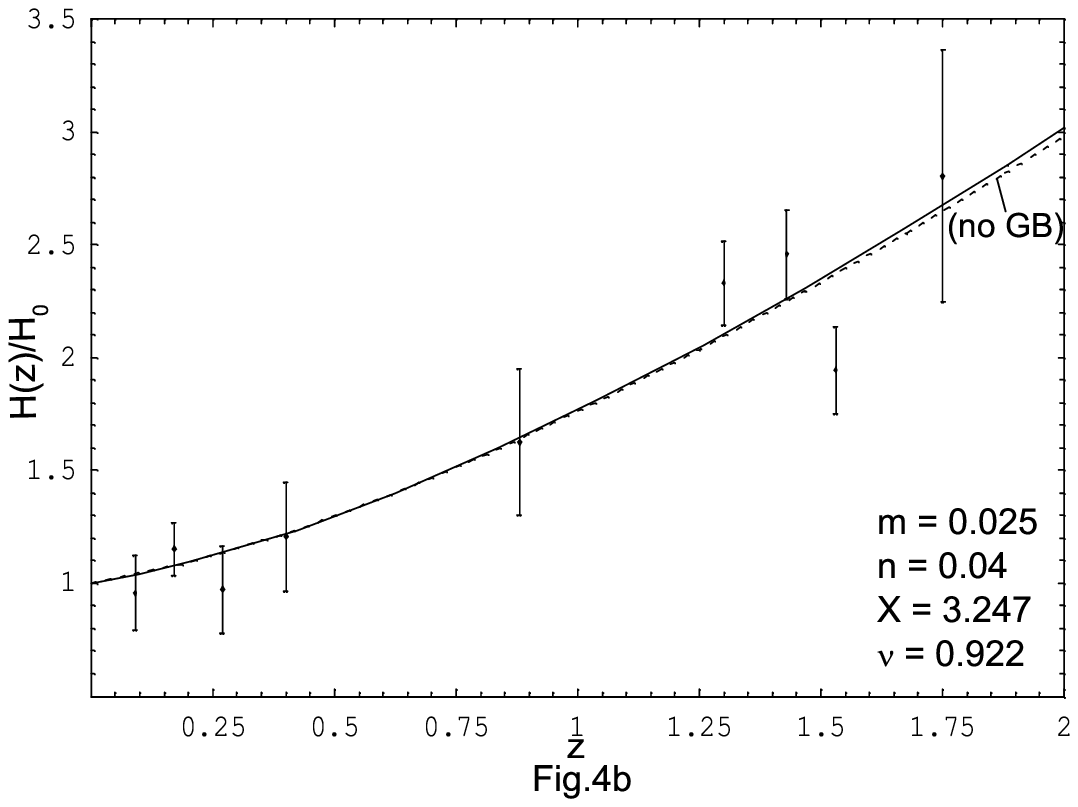}
\caption{$w_{eff}(z)$ and $q(z)$ curves (Fig.4a) and $H(z)$ curve
(Fig.4b) in DGP+GB+$T^5_5$ case. Comparisons with the results
without GB corrections have been shown. Differences from the case
without the GB correction become bigger at higher redshift.}
\label{fig4}
\end{figure}

5.DGP+$T^5_5$+M

Based on case 3, we include the dark radiation contribution. We now
have four parameters, namely $n$, $X$, $\nu$ and $M$. Employing
three constraints ($x(0)=1$, $w_{eff}(0.2)=-1$ and
$w_{eff}(0)=-1.06$), and searching $n$ in the range $0.001\leq
n\leq5$, we find that the solutions exist only when $n<0.36$. We see
that $M$ can either be positive or negative: for the negative $M$,
$w_{eff}$ never decreases with the increase of z within $z<5$; while
for the positive $M$, $w_{eff}$ drops at large $z$, and
singularities of $w_{eff}(z)$ and $q(z)$ appear within $z<5$.
Pictures of these two cases are shown in Fig.5a and Fig.5b. It is
also observed that with the increase of $n$, the positive value of
$M$ becomes bigger and the singularity appears at smaller $z$.

\begin{figure}[!h]
\includegraphics[totalheight=8cm, width=8cm]{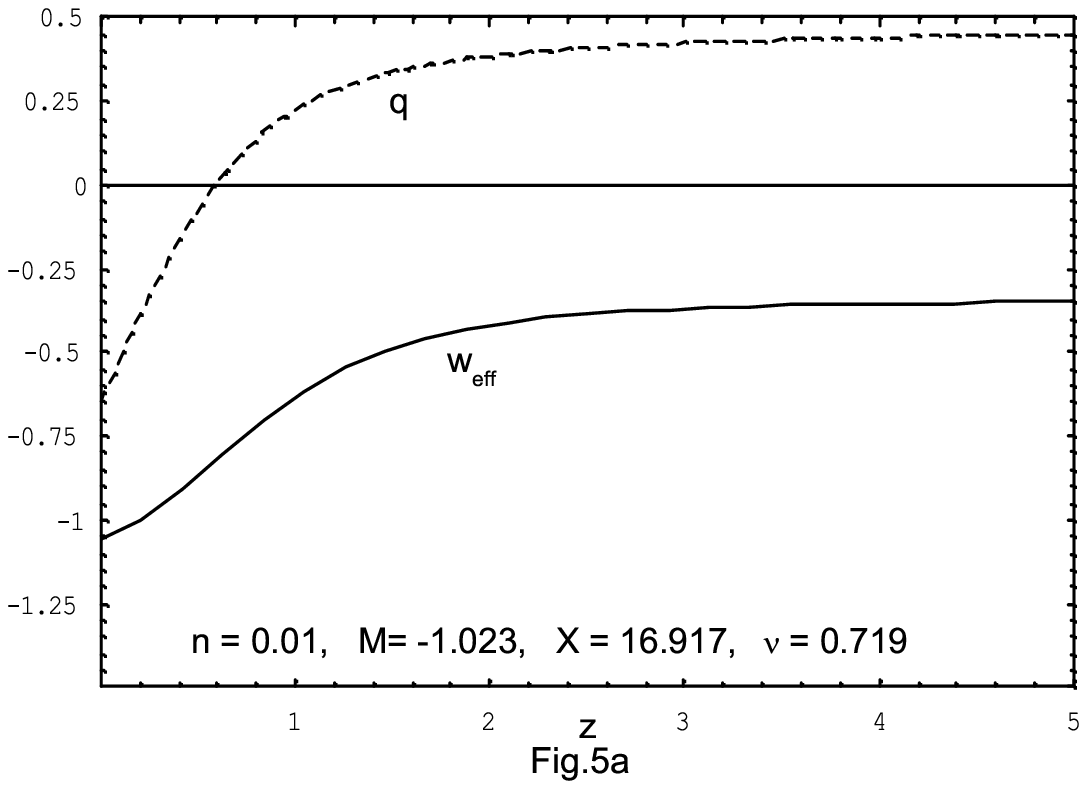}
\includegraphics[totalheight=8cm, width=8cm]{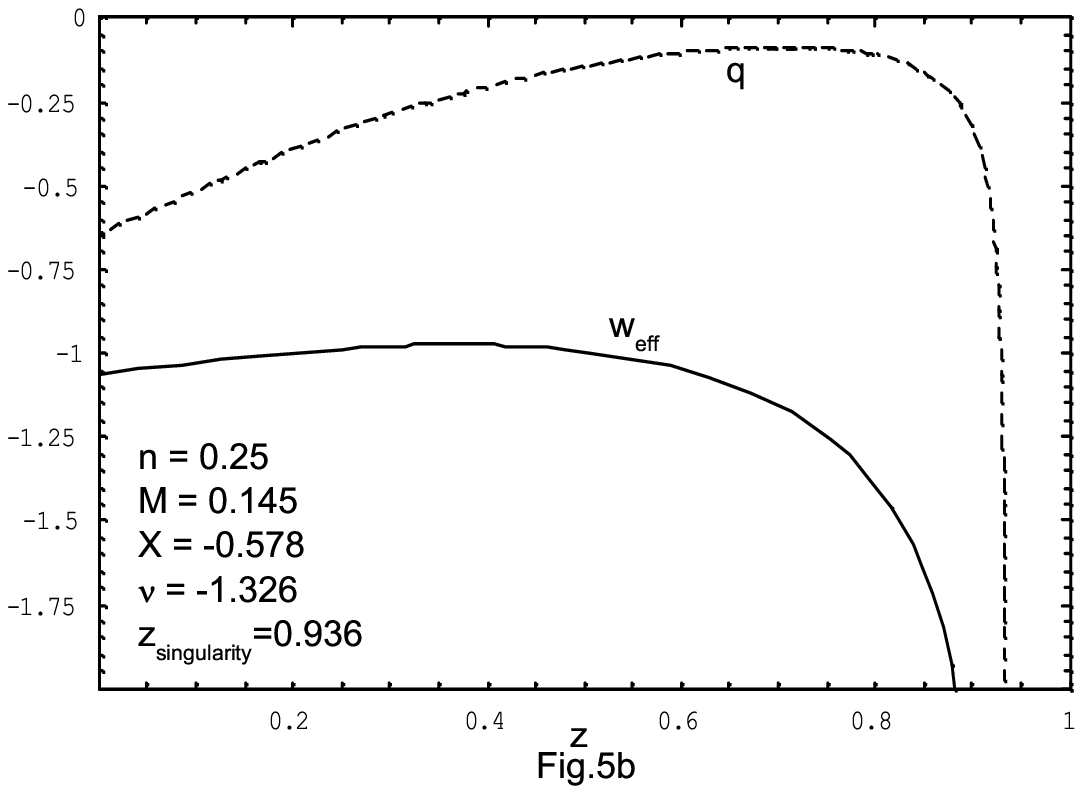}
\caption{We show in Fig.5a the $w_{eff}(z)$ and $q(z)$ curves in
DGP+$T^5_5$+M case, where M is negative and the curves are
singularity free; in Fig.5b, M is positive and the curves have a
singularity at $z=0.936$.} \label{fig5}
\end{figure}

6. DGP+GB+$T^5_5$+M

Now we have all five parameters: $n$, $m$, $X$, $\nu$ and $M$ and
three constraints to be used ($x(0)=1$, $w_{eff}(0.2)=-1$ and
$w_{eff}(0)=-1.06$). First, we try to set $m$ small, e.g.
$10^{-9}$, and choose one solution obtained in case 5, but the
\textit{FindRoot} command cannot help to get a solution to meet
all three constraints, even though the initial parameters are set
closely to those in case 5. This means that the solution is
sensitive to $\alpha$, though $\alpha$ (or $m$) is small. Next we
search the nonsingular solutions within the parameters' ranges
$0.001\leq m\leq0.02$, $0.01\leq n\leq0.1$, the curves are not
quite different from what we have seen in case 3 and case 5. We
show one of the solutions in Fig.6a and Fig.6b.

\begin{figure}[!h]
\includegraphics[totalheight=8cm, width=8cm]{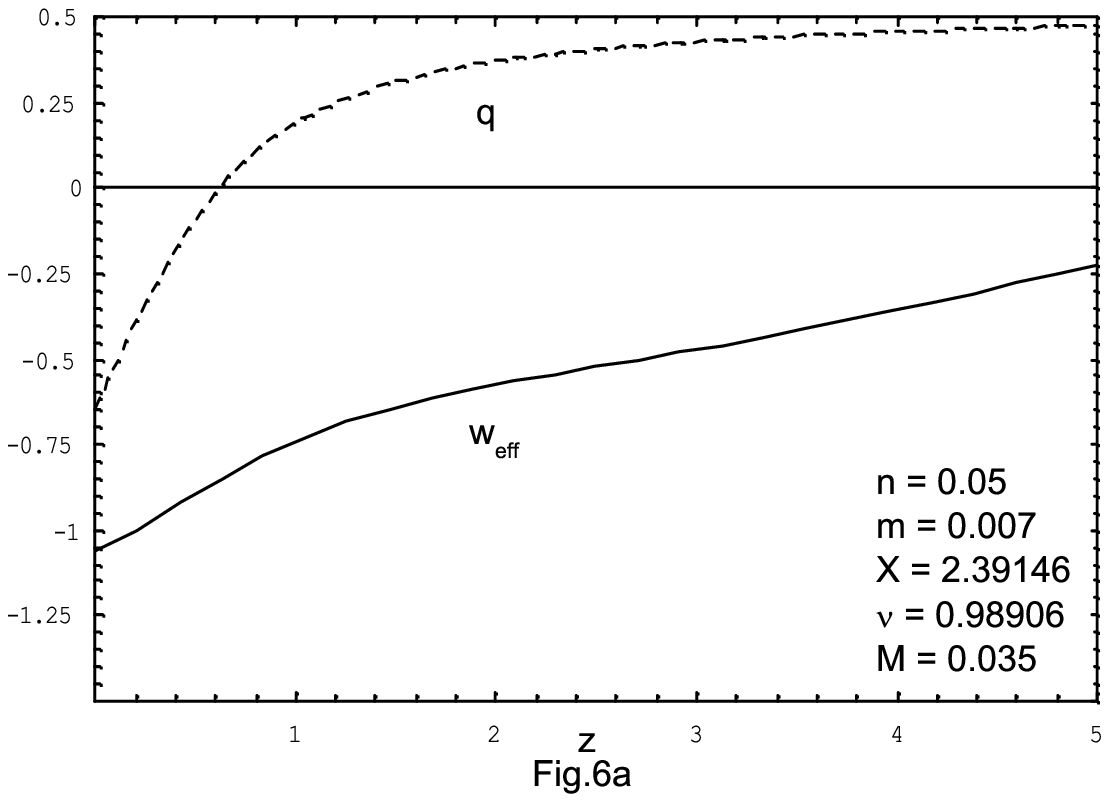}
\includegraphics[totalheight=8cm, width=8cm]{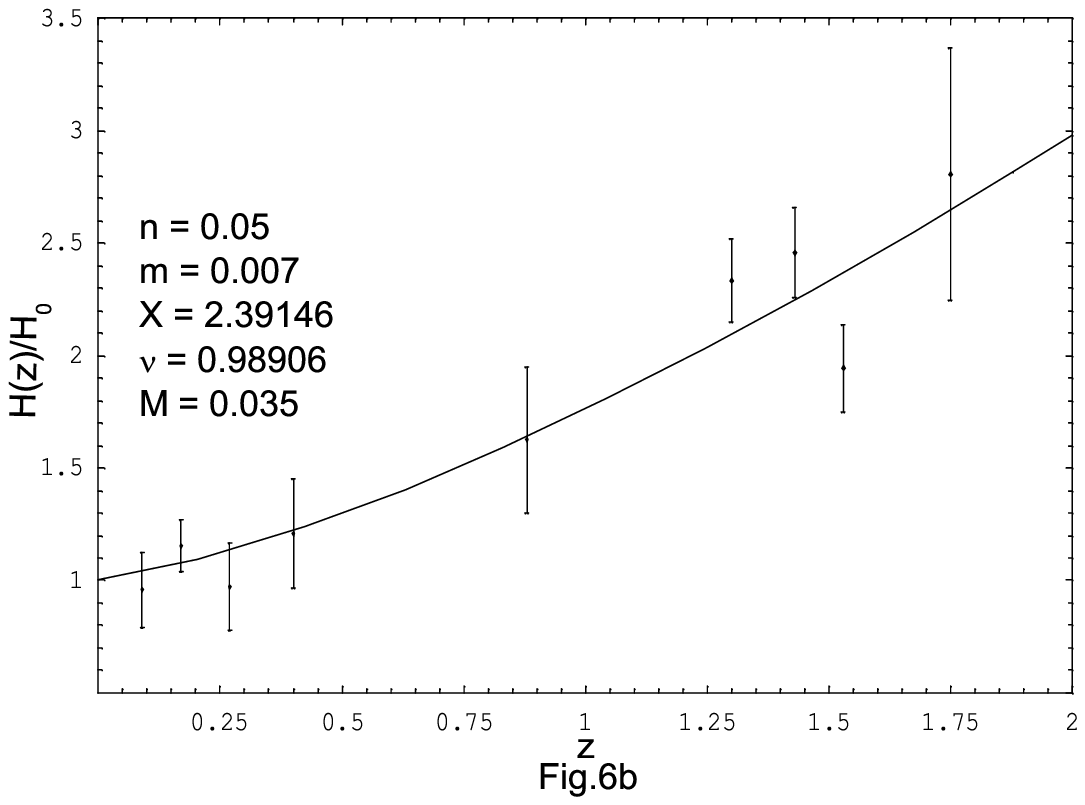}
\caption{$w_{eff}(z)$ and $q(z)$ curves (Fig.6a) and $H(z)$ curve
(Fig.6b) in DGP+GB+$T^5_5$+M case.} \label{fig6}
\end{figure}

\section{Calculation and Analysis considering the $T^0_5$ term}

The study with nonzero $T^0_5$ becomes more complicated. In order to
solve the effective Friedmann equation, or equivalently, to obtain
$x(z)$, we have to use the differential equation (23) with the full
expression of $u$ in Eq.(28). The free parameters we have now are
$n$, $m$, $\widetilde{X}$, $\nu$, $P$ and $s$. Here we do not have
the explicit dark radiation term $M$, since we do not make the
integration to get $\Phi$. In the calculation, we have to solve the
differential equation, where the boundary condition $x(z=0)=1$ is
needed. The effect of the dark radiation is reflected in the
boundary condition of $x$. Solving equations when $T^0_5=0$, the
result is the same as the case when dark radiation term $M$ appeared
in the last section. We will still use two constraints on $w_{eff}$,
which are $w_{eff}(0.2)=-1$ and $w_{eff}(0)=-1.06$.

Eq.(23) is a differential equation of $x(z)$, where free parameters
are involved. More efforts are needed to solve the equation
numerically. Here we try to employ the self-consistent method. First
we get $x(z)$ solved with the chosen initial values of parameters,
then we substitute the numerical result into the expression of
$w_{eff}(z)$. The term $\frac{dx(z)}{dz}$ in the expression of
$w_{eff}$ can be replaced by the function of $x(z)$ using Eq.(23),
thus we can write $w_{eff}$ into a function of $x(z)$. Substituting
the solution of $x(z)$, $w_{eff}$ becomes a function of $z$ and we
can solve the parameters with the constraints $w_{eff}(0.2)=-1$ and
$w_{eff}(0)=-1.06$. If the result is not consistent, we substitute
the results back to $x(z)$ as new initial values until convergence
is finally arrived. A proper choice of the initial parameters is
crucial to obtain a convergent result, we usually do the iteration
with many different choices within quite a large reasonable
parameter space.

Now we show the results we have obtained. First, to show the
consistent and efficient of our numerical calculation, we turn off
the contribution of $T^0_5$ to recover corresponding cases in
section III.

7. DGP+$T^5_5$ (employing Eq.(23) to solve the problem numerically)

We have three parameters $n$, $\widetilde{X}$ and $\nu$ in this
case. Searching within ranges $-100\leq\widetilde{X}\leq100$ and
$-4\leq\nu\leq100$ by setting $n=0.01$, we can find the solution
($\widetilde{X}=79.824$, $\nu=0.719$), which is singularity free at
least for $z<5$. This result can be compared with that in case 5
(DGP+$T^5_5$+M). Here the dimensionless parameter for $T^5_5$ term
is $\widetilde{X}$, which differs from $X$ employed in case 5 with a
factor $4+\nu$ as shown in Eq.(18). Taking this into account, the
solution for $X$ is $16.917$, which is just the result in case 5
with $n=0.01$, where the solution in case 5 reads $X=16.917$,
$\nu=0.719$ and with the additional parameter $M=-1.023$, see
Fig.5a. They coincide, although they are obtained by completely
different methods. This shows the correctness of the self-consistent
method we used and also demonstrates the equivalence of the boundary
condition in differential equation and the extra freedom of the
integration constant.

8. DGP+GB+$T^5_5$ (employing Eq.(23) to solve the problem
numerically)

We have now four parameters $n$, $m$, $\widetilde{X}$ and $\nu$.
Within ranges $-100\leq\widetilde{X}\leq100$, $0\leq\nu\leq5$ by
setting $n=0.05$ and $m=0.007$ (which are the values used in case 6
in Fig.6a and Fig.6b), we can find the solution
$\widetilde{X}=11.93120$ and $\nu=0.98898$. Considering the
difference between the dimensionless notations, this corresponds to
$X=2.39151$ and $\nu=0.98898$, which is a bit different from those
directly obtained in case 6 as $X=2.39146$, $\nu=0.98906$. This
small difference is due to the approximation we have taken in
Eq.(23), where the expansion on $\alpha$ is kept only to the linear
order. So when $\alpha\neq0$, the equation systems in case 6 and
case 8 are not exactly the same. We can see that the difference
between the curves in Fig.7a, 7b and those in Fig.6a, 6b lies in
large $z$ region, which shows that the effect of GB correction is
important in the large redshift era.

\begin{figure}[!h]
\includegraphics[totalheight=7.8cm, width=8cm]{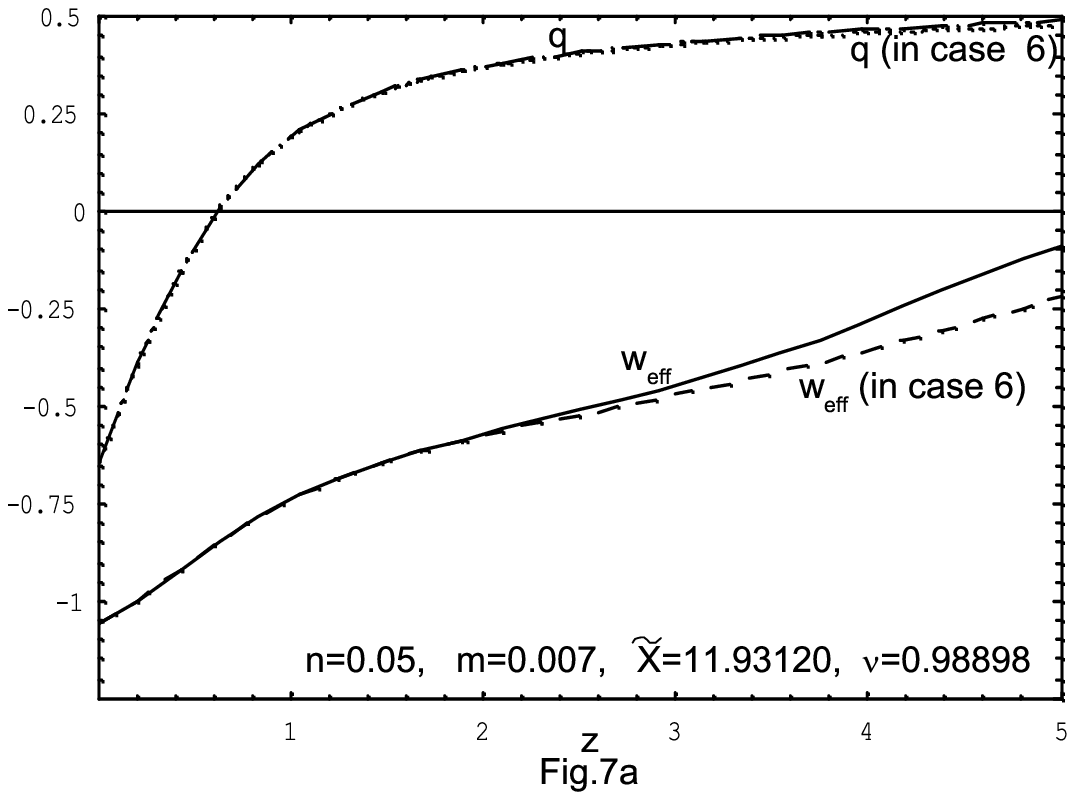}
\includegraphics[totalheight=7.8cm, width=8cm]{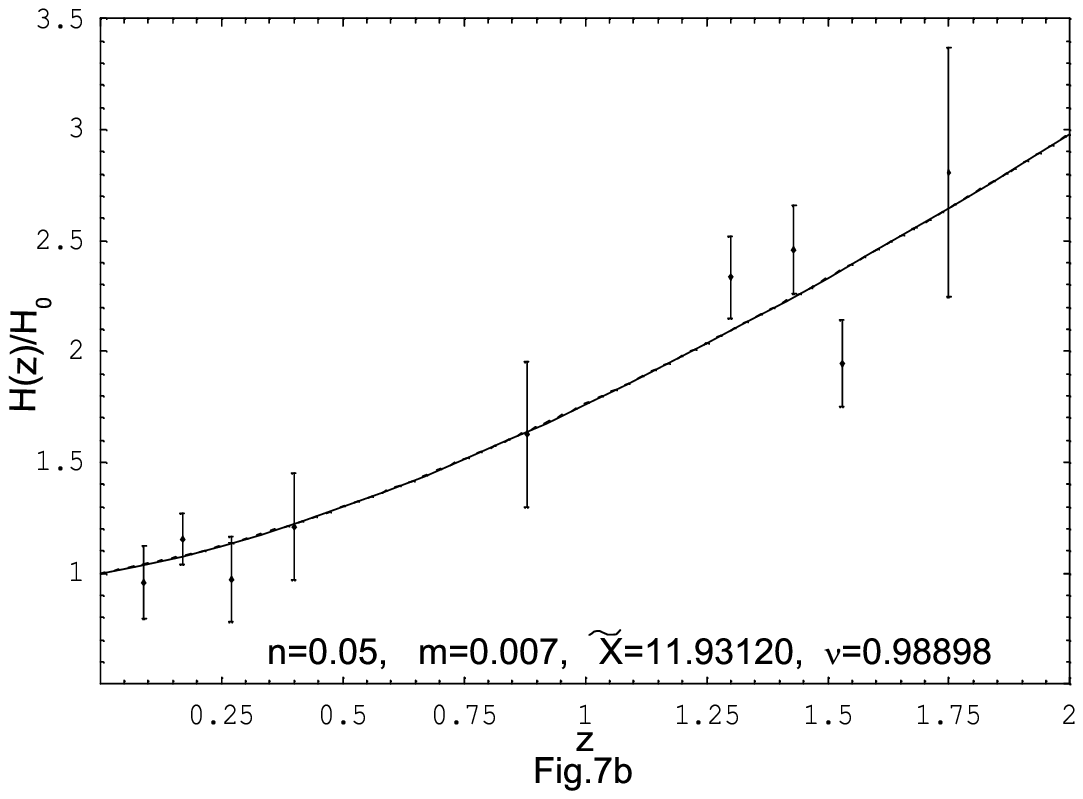}
\caption{$w_{eff}(z)$ and $q(z)$ curves (Fig.7a) and $H(z)$ curve
(Fig.7b) in DGP+GB+$T^5_5$ case. This result is calculated from the
equation considering the $T^0_5$ term, which is not quite different
from Fig.6 of case 6, except the curves lie a little higher than the
curves in Fig.6a at large $z$. } \label{fig7}
\end{figure}

To demonstrate more explicitly the effect of GB term, we calculate
in this case with different fixed values of $m$. We shut down the
freedom of $\nu$ by setting $\nu=1$ in order to show the effect of
GB term more clearly. The results are shown in Fig.8, where we see
clearly that the GB term only changes the property in the early
universe.

\begin{figure}[!h]
\includegraphics[totalheight=10cm, width=10cm]{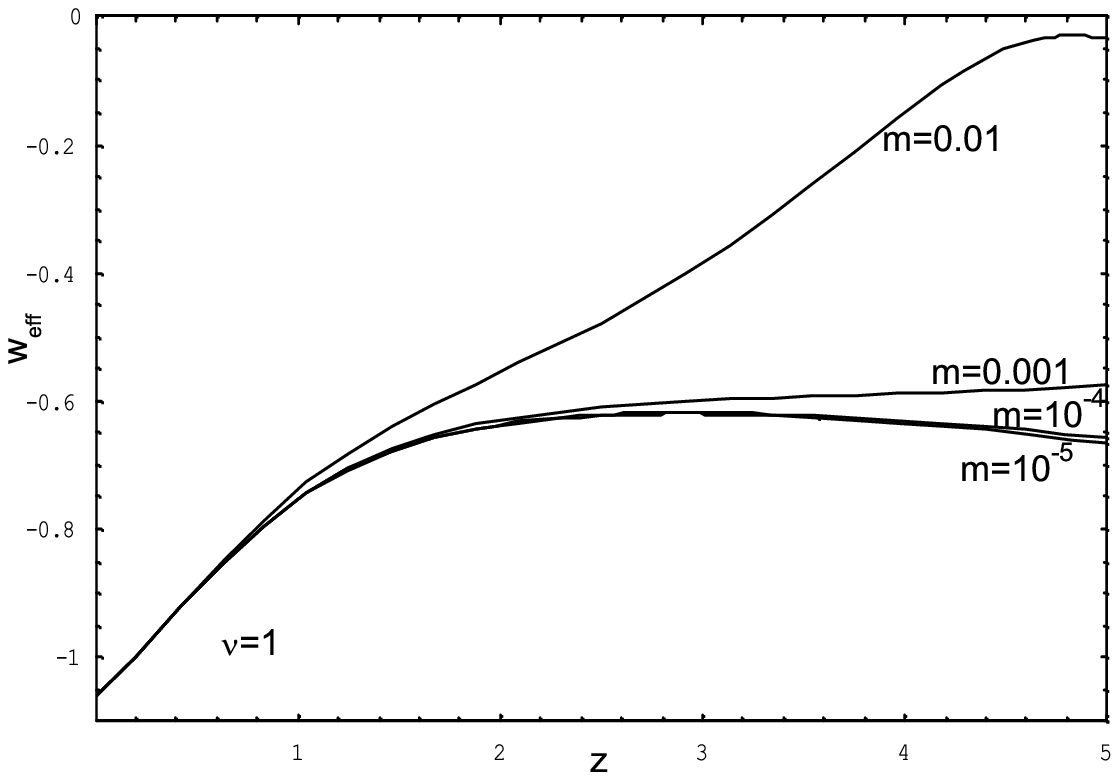}
\caption{$w_{eff}(z)$ curve in DGP+GB+$T^5_5$ case. These curves
correspond to $m=0.01$, $m=0.001$, $m=0.0001$ and $m=0.00001$
respectively. } \label{fig8}
\end{figure}

9. DGP+$T^0_5$

We turn on the $T^0_5$ effect to consider the energy exchange
between the bulk and the brane. In this case free parameters are
$n$, $P$ and $s$. Searching in the range $0.01\leq n\leq0.2$ with
$-0.28\leq P\leq0.72$ and $-100\leq s\leq100$ as initial tries, we
find that for $n\leq0.15$, solutions can be found, but all require
$P<-0.28$, which is forbidden as we discussed. This tells us that
$T^0_5$ alone with the simple form $T^0_5=fHa^s$ cannot lead to the
expected behavior of the effective equation of state.

10. DGP+GB+$T^0_5$

We have four parameters $n$, $m$, $P$ and $s$, one more than those
in the previous case. But after searching in ranges $0.001\leq
n\leq0.1$, $0.01\leq m\leq0.9$, with $-0.28\leq P\leq0.72$ and
$-100\leq s\leq100$ as initial values, all solutions which can be
found needs $P<-0.28$. The value of $P$ increases with the increase
of $m$ and the decrease of $n$, but it can only go up to $-0.425$
when $n=0.001$ and $m=0.9$, which is almost the most favored
parameter set within acceptable ranges for $m$ and $n$. Actually $m$
is related to the GB correction which should be small. Thus
introducing one more free parameter, the GB correction, cannot
change the unfavored result in case 9.

11. DGP+$T^5_5$+$T^0_5$

We contain now five parameters in total, e.g., $n$, $\widetilde{X}$,
$\nu$, $P$ and $s$, but for simplicity, we keep only three
parameters ($n$, $P$ and $\widetilde{X}$) free, by setting
$\nu=0.7$, $s=1$. The reason of choosing $\nu=0.7$ is because in
case 7 the solutions give the value of $\nu$ around $0.5\sim1$. We
find that the value of $n$ cannot be too big, otherwise the curve
will have singularity at very small $z$. For small $n$, we can find
solutions without singularity, e.g., $n=0.01$, $P=0.020$ and
$\widetilde{X}=78.535$. We show the proportion of different
components as functions of $z$ in Fig.9, in which
$\Omega_{matter}+\Omega_{dark energy}\equiv1$, $\Omega_{matter}$ is
the total matter as obtained from the differential equation (26);
while $\Omega_{effective}$ is the effective dark energy proportion
including the  energy exchange effect between the bulk and brane.
Since the energy exchange effect is very small ($P$ is small), the
difference between $\Omega_{effective}$ and $\Omega_{dark energy}$
can basically be neglected. The behavior of $w_{eff}(z)$, $q(z)$ and
$H(z)$ are shown in Fig.10a and Fig.10b. Different from the case 9,
we see that when the bulk matter $T^5_5$ is considered, the modified
DGP model allows the $w_{eff}$ crossing $-1$ and is consistent with
$H(z)$ data.

\begin{figure}[!h]
\includegraphics[totalheight=10cm, width=10cm]{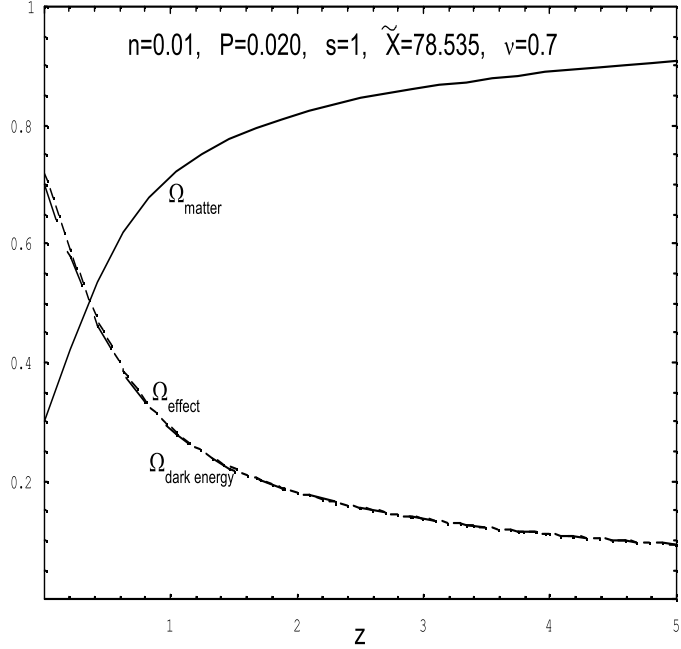}
\caption{Different components as functions of $z$ in
DGP+$T^5_5$+$T^0_5$ case. The solid line and the long dashed line,
representing the ``real" matter component and dark energy
component respectively.  The short dashed line shows the remainder
of the total energy density after subtracting the conserved matter
$\Omega_{m0}(1+z)^3H_0^2/H(z)^2$, which acts as the effective dark
energy where the energy exchange was considered. Since $P$ is
quite small, the effect of energy exchange is negligible, curves
of $\Omega_{effective}$ and $\Omega_{dark energy}$ lie almost
together. } \label{fig9}
\end{figure}

\begin{figure}[!h]
\includegraphics[totalheight=8cm, width=8cm]{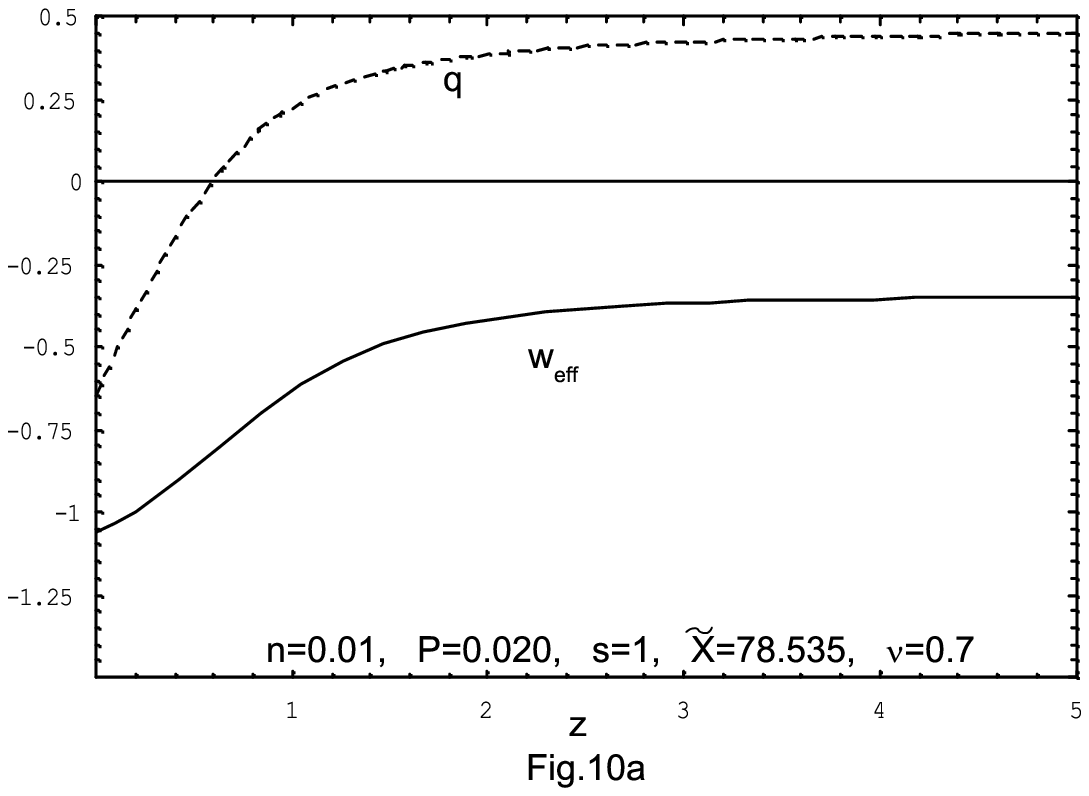}
\includegraphics[totalheight=8cm, width=8cm]{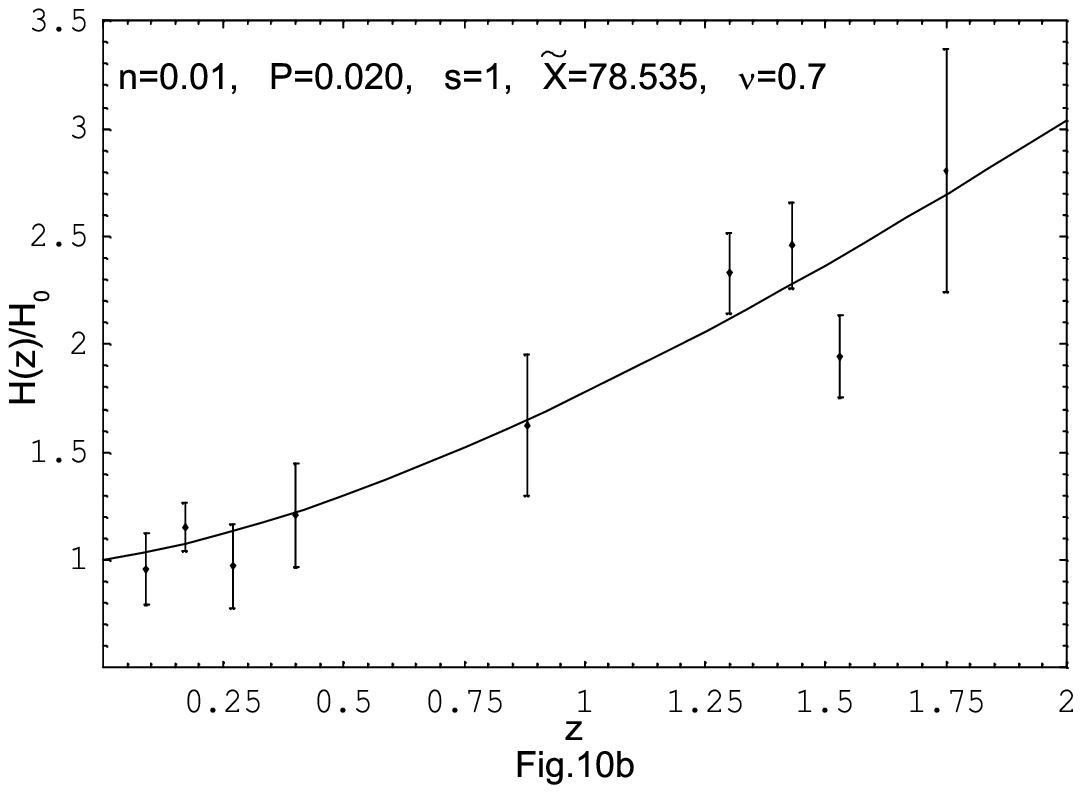}
\caption{$w_{eff}(z)$ and $q(z)$ curves (Fig.10a) and $H(z)$ curve
(Fig.10b) in DGP+$T^5_5$+$T^0_5$ case.} \label{fig10}
\end{figure}

12. DGP+GB+$T^5_5$+$T^0_5$

This is the most general case in our discussion, where we have all
parameters of our model: $n$, $m$, $\widetilde{X}$, $\nu$, $P$ and
$s$. To simplify the calculation, we fix $\nu=0.5$ and $s=1$.
Searching in parameters' ranges $0.01\leq n<10$, $0.001\leq m<1$,
$-0.28\leq P\leq0.72$ and $-1000\leq\widetilde{X}\leq1000$, we found
that when $n$ becomes larger, the solution becomes worse, either the
curves have very bad shapes or the value of $P$ lies far away from
the acceptable range. It is found that generally $n$ should not be
larger than 0.1. For small $n$, we can find the solution such as
$n=0.001$, $m=0.01$, $P=0.166$, $\widetilde{X}=895.044$, whose
corresponding curves of $w_{eff}(z)$, $q(z)$ and $H(z)$ are shown in
Fig.11a and Fig.11b respectively.

\begin{figure}[!h]
\includegraphics[totalheight=8cm, width=8cm]{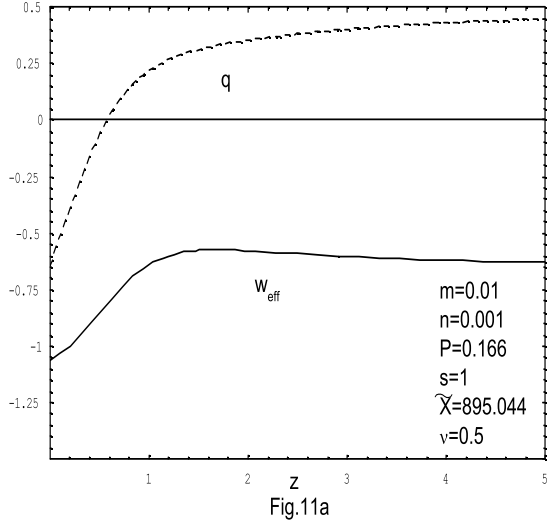}
\includegraphics[totalheight=8cm, width=8cm]{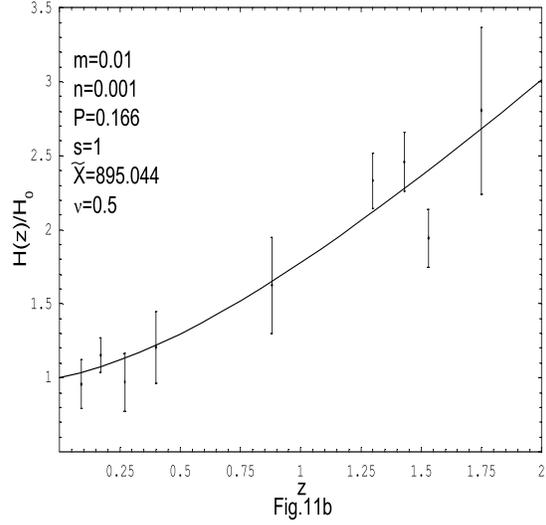}
\caption{$w_{eff}(z)$ and $q(z)$ curves (Fig.11a) and $H(z)$ curve
(Fig.11b) in DGP+GB+$T^5_5$+$T^0_5$ case.} \label{fig11}
\end{figure}

\section{Conclusions and Discussions}

In this work we have generalized the DGP braneworld by including
bulk matter content, bulk-brane energy exchange and adding the GB
curvature correction term in the bulk action. We have investigated
the effects of the bulk contents and the GB correction on the
evolution of the universe.  We have found that although the pure
DGP model cannot accommodate the transition of the equation of
state as indicated by recent observation, once the bulk matter
$T^5_5$ is considered, the modified model can accommodate the
$w_{eff}$ crossing $-1$. However this transition of the equation
of state cannot be realized by just considering bulk-brane energy
exchange or the GB effect but without the bulk matter
contribution. Thus $T^5_5$ plays the major role in the modified
DGP model to have the $w_{eff}$ crossing $-1$ behavior. The GB
term can have little influence on the late time behavior of the
universe, it gives modification to the equation of state at big
redshift. This is because of the fact that the GB correction
arises from the high energy theory, being negligible in our
present cold universe. Besides the $w_{eff}$ crossing behavior,
our model can describe the Hubble parameter consistently with
observation.

In our parameter space there is a generally favored range $n<0.1$,
which is crucial to have singularity free behavior in the equation
of state. From the definition $n=\frac{1}{r^2H_0^2}$, this range
of $n$ requires that the crossover factor obeys $r>3.16 H^{-1}_0$,
which is bigger than the lower bound just due to the GB
correction\cite{Brown:2006,He:2007}. Since $P$ and $m$ are
related, proper $P$ requires a bit bigger value of $m$. The
permitted value $P$ is small. Its sign corresponds to the
direction of the energy flow. The results we show previously in
case 11 and case 12 have positive $P$ standing for the influx of
energy, which is considered reasonable as the explanation of the
accelerating expansion of the universe for the cosmology without
extra dimension. But in the brane cosmology, since we have shown
with our results that the $T^5_5$ term dominates the effective
equation of state behavior, there is no big difference whether the
energy flows into or out of the brane, and in fact the solutions
with negative $P$ have also been obtained, which have similar
behavior to those shown here.

From our result we see that the $T^0_5$ term plays little effect in
the transition of equation of state, this could be due to the choice
of the \textit{ansatz}. A more general form of $T^0_5$ can make the
calculation more difficult, since the numerical solution rather than
the analytical solution of $\rho$ from Eq. (25) will bring more
difficulties in the following calculations.  The solution of
$\rho(a)$ should be substituted into the function of $H(z)$ after it
is expressed in dimensionless notation. But in principle this is not
impossible. Our method to numerically solve the nonlinear
differential equation of $H(z)^2$ supplies a general way to deal
with such problem, and it can relax the assumption form for $T^0_5$
and $T^5_5$. We expect to see the influence of a more general form
of $T^0_5$ on the behavior of the equation of state of effective
dark energy.

\section*{Acknowledgements}
This work is partially supported by CNPq (Conselho Nacional de
Desenvolvimento Cientifico e Tecnologico) and FAPESP (Fundacao de
Ampara a Pesquisa do Estado de Sao Paulo). The work of B. Wang was
partially supported by NNSF of China and Shanghai Education
Commission. The work of C.-Y. L. was supported in part by the
National Science Council under Grant No. NSC- 93-2112-M-259-011.

\end{document}